\documentclass{aa}
\usepackage[varg]{txfonts}
\usepackage{dcolumn}
\usepackage{natbib}
\usepackage{gensymb}
\usepackage{amssymb}
\usepackage{threeparttable}
\usepackage{graphicx}
\usepackage{tablefootnote}
\usepackage[keeplastbox]{flushend}

\renewcommand{\arraystretch}{1.2}

\begin{document}

\title{Confinement of the solar tachocline by a cyclic dynamo magnetic field} 

\author{Roxane Barnab\'e\inst{1,2}
			\thanks{Corresponding author : roxane.barnabe@cea.fr} 
  	    \and Antoine Strugarek\inst{1,2} 
        \and Paul Charbonneau\inst{2}
        \and Allan Sacha Brun\inst{1}
        \and Jean-Paul Zahn\inst{3}
             \thanks{Jean-Paul Zahn passed away while this work was underways. It is published here in his memory.} 
        } 

\institute{Laboratoire AIM, DSM/IRFU/SAp, CEA Saclay, F-91191 Gif-sur-Yvette Cedex, France
  \and D\'epartement de Physique, Universit\'e de Montr\'eal, C.P. 6128 Succ. Centre-ville, Montr\'eal, Qu\'ebec, H3C 3J7, Canada
  \and Observatoire de Paris, CNRS UMR 8102, Universit\'e Paris Diderot, 5 place Jules Janssen, 92195, Meudon, France} 

\abstract 
{The surprising thinness of the solar tachocline is still not understood with certainty today. Among the numerous possible scenarios suggested to explain its radial confinement, one hypothesis is based on Maxwell stresses that are exerted by the cyclic dynamo magnetic field of the Sun penetrating over a skin depth below the turbulent convection zone.} 
{Our goal is to assess under which conditions (turbulence level in the tachocline, strength of the dynamo-generated field, spreading mechanism) this scenario can be realized in the solar tachocline.} 
{We develop a simplified 1D model of the upper tachocline under the influence of an oscillating magnetic field imposed from above. The turbulent transport is parametrized with enhanced turbulent diffusion (or anti-diffusion) coefficients. Two main processes that thicken the tachocline are considered; either turbulent viscous spreading or radiative spreading. An extensive parameter study is carried out to establish the physical parameter regimes under which magnetic confinement of the tachocline that is due to a surface dynamo field can be realized.} 
{We have explored a large range of magnetic field amplitudes, viscosities, ohmic diffusivities and thermal diffusivities. We find that, for large but still realistic magnetic field strengths, the differential rotation can be suppressed in the upper radiative zone (and hence the tachocline confined) if weak turbulence is present (with an enhanced ohmic diffusivity of $\eta > 10^{7-8} \, cm^2/s$), even in the presence of radiative spreading.} 
{Our results show that a dynamo magnetic field can, in the presence of weak turbulence, prevent the inward burrowing of a tachocline subject to viscous diffusion or radiative spreading.} 


\maketitle

\section{Introduction}

\defcitealias{FD2001}{FDP01}
\defcitealias{Spiegel1992}{SZ92}

After several decades of observations, one of the great achievements of helioseismology remains the inversion of the internal rotational profile of the Sun \citep{Brown1989}. 
The surface differential rotation prevails through the whole convective zone \citep{Thompson2003} and, in the radiative interior, the rotation rate is uniform and matches the rotation rate of the convection envelope at mid-latitudes ($\sim 30 \degree $).
The transition from differential to uniform rotation occurs in a transition layer known as the tachocline. 
This region, which lies just beneath the convection zone, is very thin (less than $5\%$ of the solar radius, see \citealp{Charbonneau1999}) and, as a result, is subject to strong latitudinal and radial shears.

\citet[][hereafter SZ92]{Spiegel1992} developed the first hydrodynamical
model of the tachocline. 
They identified the phenomenon of radiative spreading, also known as differential rotation burrowing \citep{Haynes1991}, which was further numerically tested by \cite{Elliott1997}
and more recently confirmed with turbulent numerical experiments \citep{Wood2012}. 
By the age of the present-day
Sun, this process should have made the differential rotation burrow down to $0.4 \,R_{\odot}$. 
Hence, it has to be countered to explain the observed tachocline's thinness. 
This requires a physical mechanism that can redistribute angular momentum
latitudinally, from equatorial to polar regions,
on a timescale shorter than radiative spreading transports
it downwards.

As a first attempt to explain what stops the burrowing of the tachocline, \citetalias{Spiegel1992} proposed that the tachocline is turbulent and that horizontal turbulent transport would be enough to erase the latitudinal
differential rotation imprinted from above. 
This turbulence could be caused by penetration of convective plumes below the convective zone or shearing instabilities owing to differential rotation in the overlying envelope. 
Owing to the strongly stable stratification in the tachocline, significant
inhibition of vertical flows is expected, so that the turbulence and 
its associated transport would be strongly
anisotropic, with horizontal transport operating on much shorter
timescale than radial transport \citep{Michaud1998}.

The effect of turbulence on angular momentum
transport in the tachocline is complex and not well understood. 
Based on examples from geophysical flows \citep{Haynes1991}, \cite{Gough1998} 
claim that it should be anti-diffusive (counter-gradient) in the latitudinal direction. 
However, since the tachocline is characterized by both radial and latitudinal shears, the situation is likely more complicated and the analogy to geophysical flows must be contemplated with caution. 
Others have shown that the transport could be diffusive (down-gradient) latitudinally and anti-diffusive vertically (\citealp{Miesch2003}, \citealp{Kim2005}, \citealp{Leprovost2006}, \citealp{Kim2007}). Turbulence caused by instabilities of the rotational shear tends to lead to a diffusive transport, while turbulence driven by waves tends to have an anti-difusive influence \citep{Miesch2005}. In the latter case, the waves excited by turbulent penetrative convection are influenced both by rotation (Rossby waves) and stratification (gravity waves). What type of wave is likely to dominate in the solar tachocline is still unclear today. Hence, whether the turbulent transport is down-gradient or counter-gradient critically depends on the dominant transport processes acting in the tachocline.
Finally, \cite{Tobias2007} claim that turbulent transport would be neither diffusive nor anti-diffusive because, in the presence of magnetic fields, turbulence would not have an impact on 
turbulent angular momentum transport since Reynolds and Maxwell stresses would 
cancel each other out. Since the tachocline is likely to be the seat of dynamo field organization, it is essential to consider the role of the magnetic field. 
 
The presence of a primordial magnetic field in the radiative interior, which would stop the burrowing, is another hypothesis to explain the tachocline's confinement (\citealp{Rudiger1997}; \citealp{Gough1998}; \citealp{Garaud1999}; \citealp{Garaud2002}; \citealp{Brun2006}). 
Owing to Ferraro's law of isorotation \citep{Ferraro1937}, this field could force uniform rotation in the radiative zone if entirely confined therein, but
would imprint the latitudinal rotation of the convective envelope into the radiative
interior if connected to the envelope through the tachocline
(\citealp{MacGregor1999}; \citealp{Strugarek2011}). Confinement of such a fossil field to the radiative 
interior is thus essential, and must be sustained despite the unavoidable ohmic
diffusion, which would tend to spread it into the convective zone. 

\cite{Gough1998} proposed that a meridional circulation within the tachocline
could achieve the needed confinement of the fossil magnetic field.
Some simulations were carried out to demonstrate this process (e.g. \citealp{Brun2006}; \citealp{Strugarek2011}). 
They showed that an axisymmetric magnetic field of fossil origin in the radiative zone could not be confined and would reconnect in the convection envelope and thus would not be able to prevent the spreading of the tachocline. 
\cite{AA2013} pointed out that this negative result could be due to the
parameter regime in which these simulations were carried out. Their
(laminar) model does achieve tachocline confinement at low latitudes
in a manner akin to the \cite{Gough1998} scenario, but, as of today, no 3D turbulent simulations coupling the radiative and convection zones together have been able to achieve a global (at all latitudes) confinement of the tachocline.

Nevertheless, an hypothetical fossil magnetic field in the radiative interior is not the only option.
A cyclic dynamo-generated magnetic field is known to be present in the solar convective envelope and will certainly penetrate slightly in the upper portion of the tachocline, if only via convective overshoot.
\citet[][hereafter FDP01]{FD2001} proposed that the Lorentz force
associated with this dynamo magnetic field would lead to horizontal transfer of angular momentum from equator to poles, on a timescale sufficient to confine the tachocline. 
Owing to ohmic diffusion, this field would also diffuse inward, its amplitude decreasing with depth. A magnetic field oscillating with angular frequency
$\omega$ at the boundary of a conductor will propagate down to
the electromagnetic skin depth:

\begin{equation}
\label{eqn:skin}
H_{skin}=\sqrt{\frac{2\eta}{\omega}} ~,
\end{equation}

\noindent
where $\eta$ is the magnetic diffusivity (cf., \citealp{Garaud1999}). 
In their model, \citetalias{FD2001} imposed a constant differential rotation and an oscillatory poloidal field at the base of the convection zone. This poloidal field propagates inward owing to ohmic diffusion, with exponentially decreasing amplitude due to the skin depth effect. 
It is sheared by the differential rotation, which also propagates inward owing to viscous transport. 
The shear generates a toroidal field that oscillates with the same period as the poloidal field. 
\citetalias{FD2001} showed that, under their model setup,
the Lorentz force associated by this poloidal$+$toroidal field would be able to suppress the differential rotation at the base of the magnetic layer. 
Their conclusion however comes with two important caveats: (1) the zero-toroidal
field at the upper boundary of the tachocline is hard to reconcile with current
dynamo models of the solar cycle, and (2) more importantly, their model
omits the effect of radiative spreading, which is expected to dominate
over purely viscous spreading under solar interior conditions (e.g. Small $Pr$, as stated in \citetalias{Spiegel1992}).

The primary purpose of the work presented in this paper is to address these
two limitations of the \citetalias{FD2001} model.
We thus extend their model to accomodate a more realistic upper boundary
condition on the dynamo magnetic field, namely oscillatory poloidal and toroidal
components, with a set phase lag between the two. 
We also introduce radiative spreading into the model, following the hyperdiffusion formulation of 
\citetalias{Spiegel1992},  to assess the robustness of the \citetalias{FD2001} proposal
under varying scenarios for the turbulent angular momentum transport within
the tachocline: viscous-like, anti-diffusive, or mediated by radiative spreading.
In Sect. \ref{sec:magtach}, we consider viscous-like cases (diffusive and anti-diffusive), while in Sect. 
\ref{sec:hyper}, we take the effect of radiative spreading into account. 
We conclude in Sect. \ref{sec:conclu} by summarizing our most salient
results and discussing them in the context of past and ongoing
numerical simulations of the tachocline.

\section{Magnetic tachocline}
\label{sec:magtach}

We design a simplified tachocline model to test the validity of \citetalias{FD2001}'s scenario. 
We develop the model in Sect. \ref{subsec:eq}, justify our choices of representative parameters for the tachocline in Sect. \ref{subsec:param} and present our reference viscous solar model in Sect. \ref{subsec:results}. We then discuss the effect of the magnetic Prandtl number in Sect. \ref{subsec:prm} and compare cases with diffusive and anti-diffusive turbulent viscosity in Sect. \ref{subsec:diff_vs_antidif}.
	
\subsection{Equations}
\label{subsec:eq}

We consider an oscillatory magnetic field penetrating the tachocline through the bottom of the convection zone. 
We aim to characterize its ability to suppress the differential rotation sustained by the rotating turbulent convection zone, which here is supposed to extend into the deeper radiative zone. 
To simplify the problem, we use cartesian coordinates, with the azimuthal and latitudinal coordinates of the spherical system mapped onto $x$ and $y$ respectively. The vertical coordinate $z$ runs downwards.

We consider a two-components magnetic field ($B_y$ and $B_x$
being the poloidal and toroidal components, respectively),
oscillating at a frequency $\omega$ at the base of the convective zone. 
We set a constant amplitude ratio between these two magnetic field
components at the upper boundary, measured by the parameter $\beta$.
Hence, the amplitude of the poloidal field is given by $B_0$, while the toroidal field amplitude is $\beta B_0$. We consider a latitudinal differential rotation represented by a sheared velocity field $U_x$, whose amplitude and
spatial latitudinal scale are given by $U_0$ and $1/k$. We assume that the toroidal magnetic field also varies on this characteristic spatial scale. The poloidal component is consider to be constant with latitude, to simplify the problem and for the divergence of the magnetic field to be zero. All fields are assumed to be invariant along the longitudinal direction $x$ (i.e. axisymmetry). 

We introduced the dimensionless fields $a$, $b$, and $u$, respectively representing the magnetic poloidal and toroidal fields and the velocity field: 

\begin{equation}
\left\{
\begin{split}
\begin{gathered}
\label{eqn:eqlabel2}
\begin{aligned}
& B_y(z,t)=B_0a(z,t) \\
& B_x(y,z,t)=\beta B_0b(z,t)\sin(ky) \qquad  ~.\\
& U_x(y,z,t)=U_0u(z,t)\cos(ky)
\end{aligned}
\end{gathered}
\end{split}
\right.
\end{equation}

\noindent
We consider the induction equation 

\begin{equation}
\frac{\partial{\textbf{B}}}{\partial{t}}=\nabla \times \big( \textbf{u} \times \textbf{B} - \eta \nabla \times \textbf{B} \big) ~,
\end{equation}

\noindent
and the Navier-Stokes equation

\begin{equation}
\label{eqn:NS}
\frac{\partial{\textbf{u}}}{\partial{t}} + \big( \textbf{u} \cdot \nabla \big) \textbf{u} = \nu  \nabla ^2 \textbf{u} + \frac{1}{\rho c}\textbf{J} \times \textbf{B} ~, 
\end{equation}

\noindent
where $\eta$ is the ohmic diffusivity, $\nu$ is the viscosity, $\rho$ is the density of the environment and $c$ is the speed of light. We assumed hydrostatic equilibrium, so that the pressure gradient and gravity cancel out in Eq. \ref{eqn:NS}. The non-dimensional functions $a$, $b$, and $u$ thus obey the following parabolic system: 

\begin{equation}
\label{system}
\left\{
\begin{split}
\begin{gathered}
\begin{aligned}
& \frac{\partial{a}}{\partial{\tau}}=\frac{\partial^2{a}}{\partial{\xi}^2} \\
& \frac{\partial{b}}{\partial{\tau}}=\frac{\partial^2{b}}{\partial{\xi}^2}-(k\delta)^2b-C_Aau \qquad \qquad ~.\\
& \frac{\partial{u}}{\partial{\tau}}=\bigg(\frac{\nu}{\eta}\bigg)\bigg[\frac{\partial^2{u}}{\partial{\xi}^2}-(k\delta)^2u\bigg]+C_Lab 
\end{aligned}
\end{gathered}
\end{split}
\right.
\end{equation}

\noindent
Terms from other scales, which are not considered in our decomposition, are modeled through simple diffusion operators with the increased associated coefficients of viscosity $\nu$ and ohmic diffusivity $\eta$. For the time being, we assume that differential rotation is spreading inward by viscous transport. The independent variables have been non-dimensionalized by the depth of the computational domain $\delta$ and by the ohmic diffusion time $\Theta$, defined as follows: 

\begin{equation}
z=\delta\xi, 
\qquad
t=\Theta \tau, 
\qquad
\textrm{with} \; \Theta=\delta^2/\eta ~.
\end{equation}

\noindent
The dimensionless coupling constants $C_A$ and $C_L$ measure the magnitude
of induction and the Lorentz force, respectively. These are given by:

\begin{equation}
C_A=\frac{R_m}{\beta}
\quad ~; \quad
C_L=R_m\beta\Lambda ~,
\end{equation}

\noindent
where we have introduced the magnetic Reynolds number $R_m$ and the Elsasser number $\Lambda$:

\begin{equation}
R_m=k\Theta U_0=\frac{k\delta^2U_0}{\eta} 
\quad ~; \quad
\Lambda=\frac{B_0^2}{\rho U_0^2} ~.
\end{equation}

\noindent
At the top of the tachocline ($\xi=0$), we impose a poloidal and a toroidal magnetic field oscillating periodically ($\Omega=\omega\Theta$), with a phase
lag given by $\phi$. The differential rotation is considered
imposed from above by convection, leading to the following upper boundary conditions:

\begin{equation}
\left\{
\begin{split}
\begin{gathered}
\label{eqn:cond_top}
\begin{aligned}
& a(\xi=0,\tau)=\cos(\Omega \tau)  \\
& b(\xi=0,\tau)=\cos(\Omega \tau-\phi) \qquad ~. \\
& u(\xi=0,\tau)=1
\end{aligned}
\end{gathered}
\end{split}
\right.
\end{equation}

\noindent
At the bottom of our solution domain, deeper in the radiative zone ($\xi=1$), we set the three fields to zero:

\begin{equation}
\left\{
\begin{split}
\begin{gathered}
\label{eqn:cond_bot}
\begin{aligned}
& a(\xi=1,\tau)=0  \\
& b(\xi=1,\tau)=0  \qquad ~. \\
& u(\xi=1,\tau)=0
\end{aligned} 
\end{gathered}
\end{split}
\right.
\end{equation}

\noindent
Since in Sect. \ref{sec:magtach} we are interested in a tachocline propagating inward as a result of viscous momentum transport, Eqs. \ref{system}
do not yet take into account radiative spreading, the inclusion of which is
deferred until Sect. \ref{sec:hyper}.

We used a Crank-Nicolson method to solve the non-linearly coupled system of Eqs. \ref{system}. 
Although this method is unconditionally stable, numerical accuracy drops when the ratio of temporal and spatial grid spacing becomes large. This means that a smaller spatial grid size forces us to use smaller time steps. 
In order to lower the numerical cost, we implemented a non-uniform stretching in the $z$ direction, allowing us to get a higher density of mesh points in the upper part of the tachocline, just below the convection zone. 

\subsection{Choice of parameters}
\label{subsec:param}

In Table \ref{table_sect2}, we present the parameter values used in the model. 
The size of the computational domain is chosen so as to comfortably accommodate the present-day solar tachocline. 
We set $\delta=R/10$, where $R$ is the radius at $\xi=0$. In this case, we take $R=0.718 \, R_{\odot}$. 
Hence the spatial domain spans $r=0.646 \, R_{\odot}$ to $r=0.718 \, R_{\odot}$, corresponding to about $7\,\%$ of the solar radius. The horizontal wavenumber is taken as $k=2.5/R$ and therefore $k\delta=1/4$. This corresponds to the dominant scale of the solar latitudinal differential rotation in the convection zone (e.g. \citealp{Thompson2003}). For the magnetic cycle circular frequency, we take $\omega=2\pi/P$, with a period of $P=22$ years. We consider a density of $\rho=0.21\,g/cm^3$ at the base of the convection zone taken from standard solar models (e.g. \citealp{Brun2002}). The amplitude of the differential rotation is taken as $U_0=3\times 10^4\,cm/s$ \citep{Schou1998}.

The magnetic field amplitude within the tachocline is not well constrained by
observations or helioseismology (\cite{Antia2000} put an upper limit of about $300\,kG$). Models of the buoyant destabilisation and rise of thin flux tubes through the convective zone imply a toroidal magnetic field at the base of the convective envelope of the order of $10^{4-5} \, G$ (\citealp{Jouve2009}; \citealp{Fan2009}). Hence, in our model, we considered toroidal magnetic field 
amplitudes ranging from $10^3 \, G$ to $10^6 \,G$, with a preferred so-called solar value of $13750\,G$.
Most dynamo models of the solar cycle predict a poloidal-to-toroidal field
strength ratio below unity, so that we consequently adopt
$\beta=10$, which leads to a poloidal magnetic field amplitude of $1375\,G$
for our solar model, ranging from $10^2 \, G$ to $10^5 \,G$ for the purpose
of the parameter study that we report on below.

The effect of turbulence and its associated angular momentum and magnetic flux
transport in the tachocline is modeled via a linear diffusive operator
with enhanced transport coefficients. Since there are few hard constraints
on this parameter, we explore a wide range of values in the calculations
to follow, but consider that the (turbulent) magnetic Prandtl number
is set to $\rm{Pr_m}=\nu/\eta=1$.

\begin{table}
\centering
\caption{Numerical values for the parameters of the viscous solar model.}
\begin{tabular}{ccc}
\hline
\hline
 \footnotesize{Parameter} & \footnotesize{Symbol} & \footnotesize{Numerical value} 
 \\[0.5ex]

\hline
\footnotesize{Size of the domain}   				& \footnotesize{$\delta$}  			& \footnotesize{$R/10$} 					\\
\footnotesize{Radius at  $\xi=0$}					& \footnotesize{$R$}			 		& \footnotesize{$0.718 R_{\odot}$} \\
\footnotesize{Horizontal wavelength} 				& \footnotesize{$k$}      			& \footnotesize{$2.5/R$} 				\\
\footnotesize{Period} 								& \footnotesize{$P$}      			& \footnotesize{$22\,yr$} 				\\
\footnotesize{Density} 								& \footnotesize{$\rho$}   			& \footnotesize{$0.21\,g/cm^3$} 			\\
\footnotesize{Ohmic diffusivity} 				    & \footnotesize{$\eta$}			 	& \footnotesize{$10^{5}-10^{10}\,cm^2/s$}  \\
\footnotesize{Viscosity}			 					& \footnotesize{$\nu$}			 	& \footnotesize{$10^5-10^{12}\,cm^2/s$}  \\
\footnotesize{Magnetic Prandtl number} 				& \footnotesize{$\rm{Pr_m}$} 	       		 & \footnotesize{$1$} 					\\
\footnotesize{Differential rotation amplitude} 		& \footnotesize{$U_0$}    			& \footnotesize{$3\times 10^4\,cm/s$} 	\\
\footnotesize{Poloidal magnetic field amplitude} 		& \footnotesize{$B_0$}    			& \footnotesize{$10^2-10^5 \,G$} 				\\
\footnotesize{Toroidal magnetic field amplitude} 		& \footnotesize{$\beta B_0$}  		& \footnotesize{$10^3-10^6 \,G$} 	\\			
 \hline
\end{tabular}
\label{table_sect2}
\end{table}

\subsection{Reference viscous solar model}
\label{subsec:results}

\begin{figure}[h]
\centering
\includegraphics[width=0.99\linewidth]{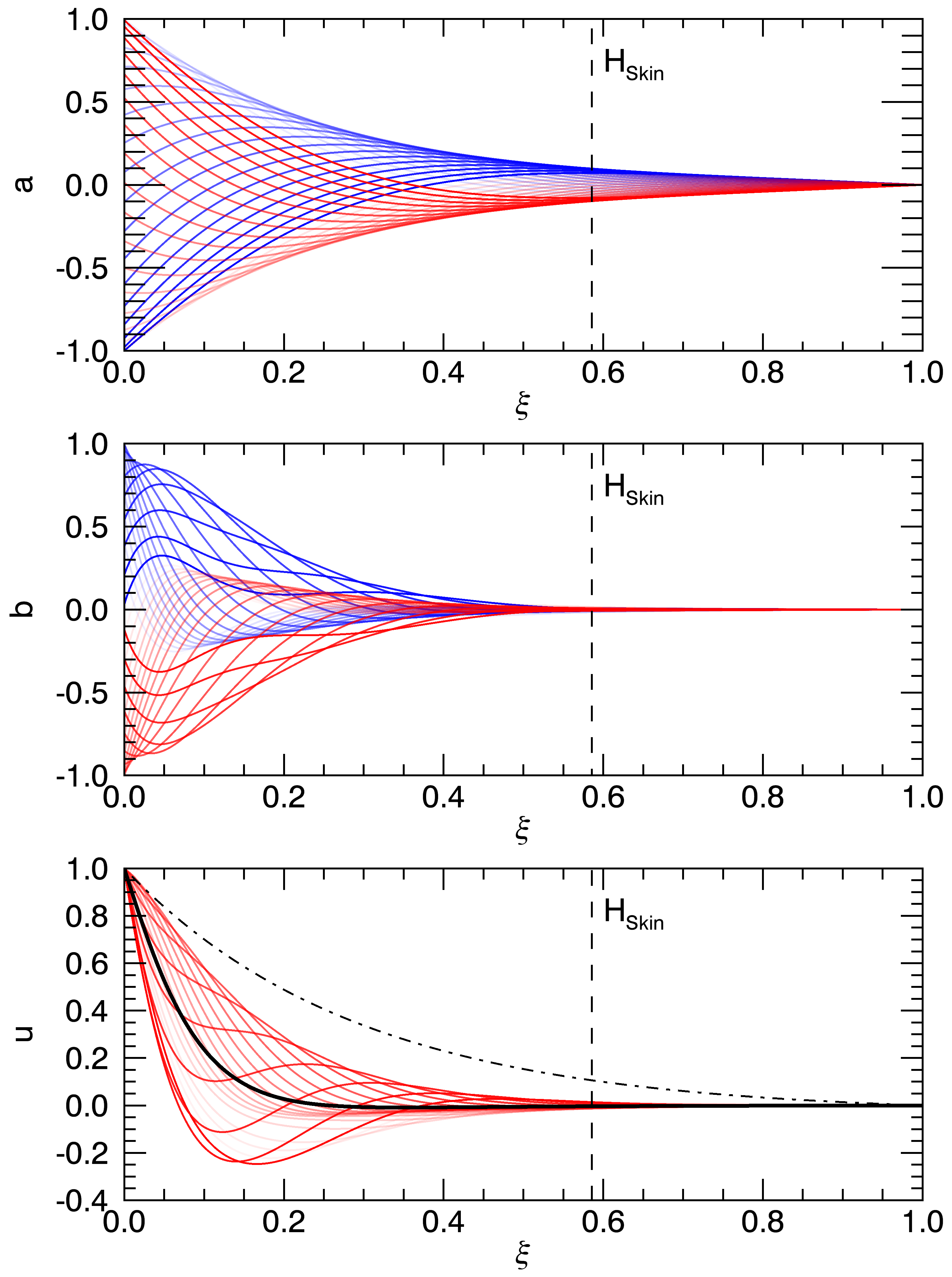}
\caption[Temporal evolution of $B_y$, $B_x$ and $U_x$]{\label{evol_visc}  \footnotesize{ (A) Temporal evolution of the poloidal field ($a$), (B) the toroidal field ($b$) and (C) the azimuthal velocity field ($u$) as a function of depth, over one magnetic cycle, for a fully relaxed simulation ($\tau=100$). Blue lines correspond to the first half of the cycle, while red lines are for the second half. For the velocity field, only half a cycle is shown as profiles are superimposed for the second half.
Lines become darker as time increases. $\xi=0$ is the top of the tachocline ($R=0.718\,R_{\odot}$) and $\xi=1$, the bottom of the domain ($R=0.646\,R_{\odot}$). 
The black line is the azimuthal velocity amplitude averaged over one cycle.
The skin depth (dashed line and Eq. \ref{eqn:skin}) and the initial condition for $u$ (dash-dotted line) are shown. 
Model parameters values for this solution are listed in table \ref{table_sect2} and the phase shift between both magnetic components is set at $\phi=\pi/2$. } }
\end{figure}

Magnetic fields oscillating at the top of the solar radiative zone will undergo
diffusive penetration owing to the electromagnetic skin effect. This is illustrated
in Fig. \ref{evol_visc}A, showing the temporal evolution of the poloidal magnetic field over one cycle for a converged simulation. 
The overall amplitude of the oscillating poloidal field, which is subject only to ohmic diffusion, is maximal at the top of the tachocline (where it is set by our
adopted boundary condition, see Eqs. \ref{eqn:cond_top})
and decreases exponentially with depth.

The toroidal component of the magnetic field is affected by both ohmic diffusion and by induction associated with the shearing of the
poloidal magnetic field by differential rotation.
Induction operates in conjunction with the skin depth effect, leading to a more complex depth variation of the toroidal magnetic component, 
as seen in Fig. \ref{evol_visc}B. 
In the case shown here, the fields are in quadrature ($\phi=\pi/2$) with the toroidal field lagging the poroidal component.

Fig. \ref{evol_visc}C, shows the evolution of the 
azimuthal velocity over one cycle in a converged simulation. We note that the quantity $k u$ is a measure of the latitudinal shear at the corresponding depth. Equally, $u$ itself is the projection of the full velocity field on one particular scale. It evolves under the joint action of viscosity and the Lorentz force, at a frequency twice that of the imposed magnetic field, owing to the quadratic dependence of the Lorentz force with the magnetic field. The azimuthal velocity oscillates about a mean value decreasing to zero with depth, yet the
latitudinal shear reverses its sign locally at some phases of the cycle,
in response to the time-varying Lorentz force.
Nonetheless, here the latitudinal
shear decreases rapidly with $\xi$ and falls below $1\%$ of its top value already at $\xi\approx 0.65$ (about $0.67\,R_{\odot}$), which means that the dynamo field is able to suppress 
the latitudinal differential rotation in the upper radiative zone, 
and thus confine the tachocline. 

This simulation reaches a stationary state very quickly, after a few diffusion times. We run the simulations for $100$ diffusion times (about $500$ magnetic cycles for $\eta=7\times 10^9 \, cm^2/s$) and find that the cycle-averaged shear profile does not evolve anymore in time and hence has reached a stationary state.

\begin{figure}[h]
\centering
\includegraphics[width=0.99\linewidth]{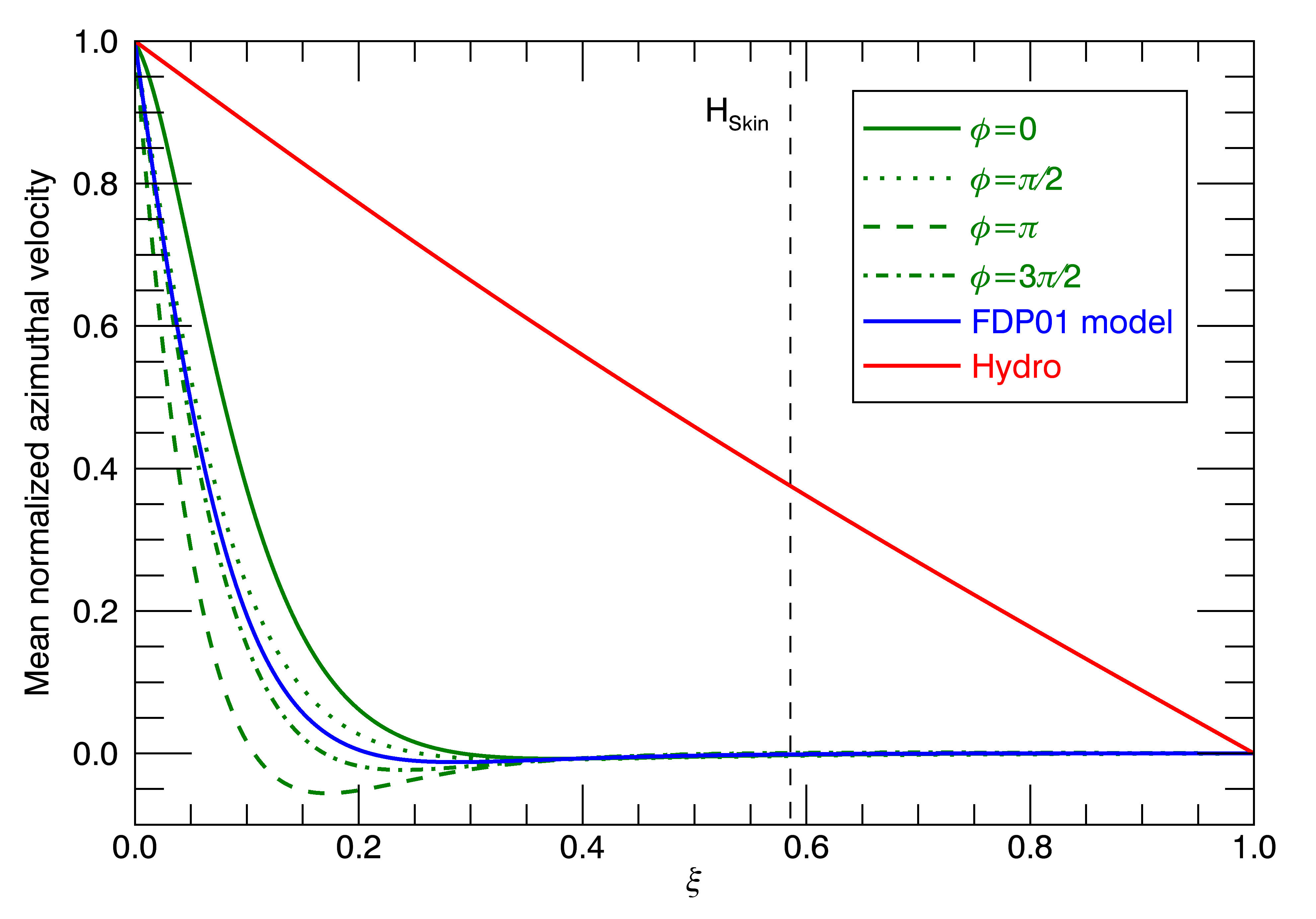}
\caption[Azimuthal velocity for different phase lags]{\label{ux_moy}  \footnotesize{Normalized azimuthal velocity averaged over one cycle as a function of depth. The red line corresponds to an hydrodynamical viscous steady state
without magnetic field. The blue curve shows results with the toroidal
component set to zero at the top of the tachocline, as
in \citetalias{FD2001}. The green lines are cases with oscillating
poloidal and toroidal magnetic fields, with different phase lags. 
Corresponding model parameter values are listed in table \ref{table_sect2}.} }
\end{figure}

We tested different phase lags $\phi$ between poloidal and toroidal magnetic components, to assess the impact of this model parameter
on the evolving latitudinal shear. Specifically, we computed solutions similar
to that plotted in Fig.~\ref{evol_visc}, but for poloidal and
toroidal magnetic components
in phase ($\phi=0$), in quadrature ($\phi=\pi/2$ or $\phi=3\pi/2$) 
and in opposition of phase ($\phi=\pi$). 
Fig. \ref{ux_moy} shows the mean profile of azimuthal velocity $u$ for these different phase lags $\phi$ (green lines). 
Differences do materialize, with the
$\phi=0$ case leading to a greater mean velocity over the solution domain, 
while the $\phi=\pi$ case shows the most pronounced reversal
within the tachocline. 
This is not surprising since, when the fields are in phase, the Lorentz force will always be positive at the top of the tachocline while, when they are in opposition, it is negative. 
As for the cases in quadrature, the Lorentz force will, at certain times, help to confine, while at other times in the cycle, it facilitates the spreading of the differential rotation. 
The nonlinear interaction due to induction mitigates this aspect and causes an asymmetry, which favors the confinement. 
Nonetheless, differences between the shear profiles for these various
phase lags remain slight and, in all cases, the vertical
spreading of the shear is stopped long before it reaches the bottom of the domain: in fact, it is kept confined well inside the skin depth of the oscillating dynamo field (vertical dashed line). This implies that the phase lag has no significant impact on tachocline
confinement under this model setup. Hence, for the remainder of our study and given solar observations of field reversal during sunspot maximum \citep{Hathaway2015}, we adopt a toroidal component in quadrature with the poloidal component, the latter
leading the former by $\phi=\pi/2$. 

\begin{figure}[h]
\centering
\includegraphics[width=0.99\linewidth]{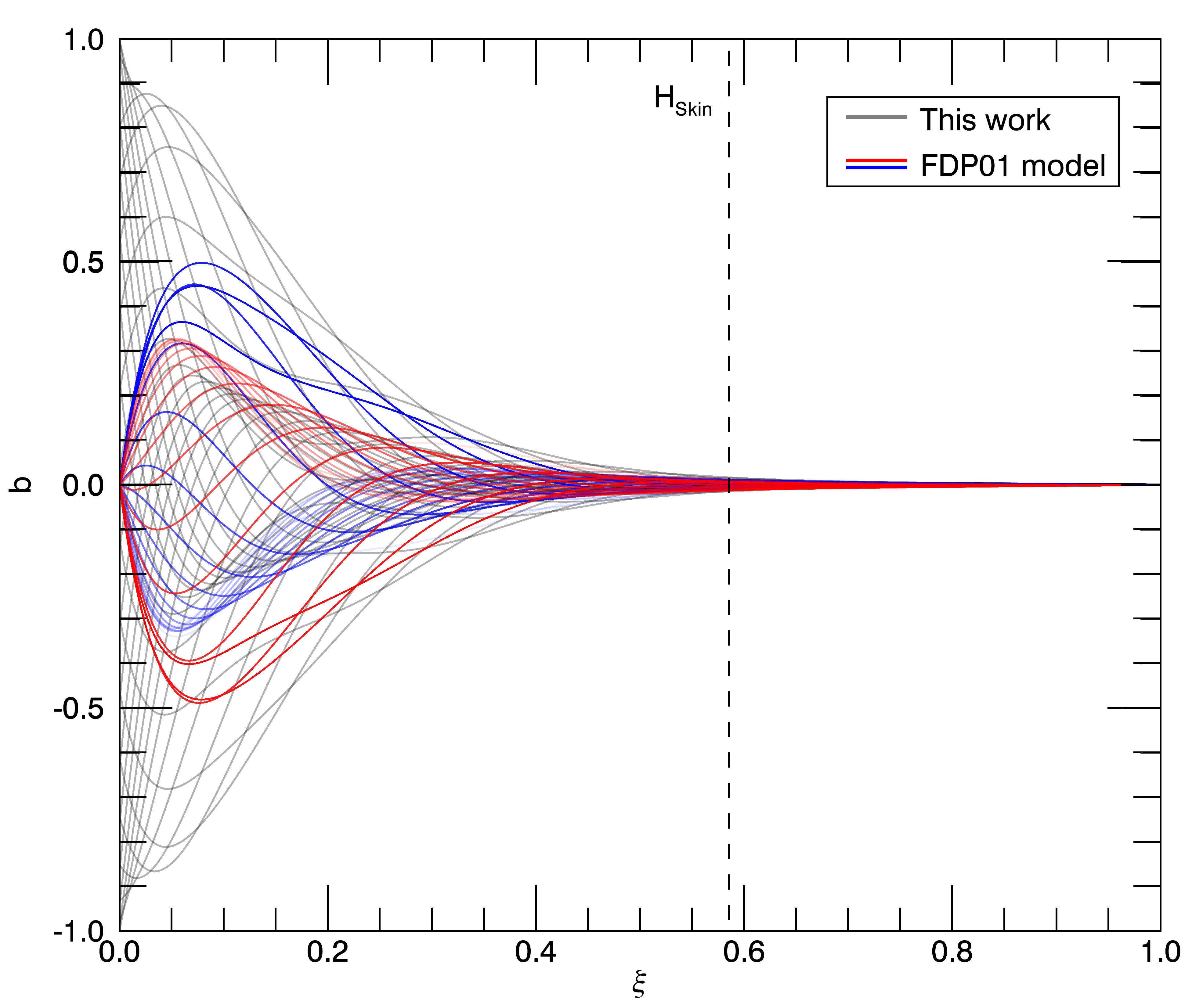}
\caption[Toroidal magnetic field]{\label{btor} \footnotesize{Temporal evolution of the toroidal magnetic field ($b$), over one cycle, for a fully relaxed simulation ($\tau=100$), with our boundary condition (grey lines) and \citetalias{FD2001} boundary condition with $\phi=\pi/2$ (in blue for the first half of the cycle, then in red, lines become darker as time evolve). The skin depth (dashed line and Eq. \ref{eqn:skin}) is also shown. 
Parameter values are listed in table \ref{table_sect2}.} }
\end{figure}

By way of comparison, a case was run using the \citetalias{FD2001} boundary condition, namely zero toroidal field at the top of the tachocline ($B_x(\xi,\tau)=0$), and is shown in blue in Fig. \ref{ux_moy}. Owing to induction, the toroidal magnetic field is still able to reach an amplitude of about $50\%$ that of the applied poloidal field (see Fig. \ref{btor}). Hence, there is no qualitative difference on the inward spreading of the differential rotation, which highlights the robustness of the confinement mechanism. In contrast, in the hydrodynamical case evolving under the sole influence of (enhanced) viscous angular momentum transport, the latitudinal shear propagates all the way
to the base of the domain, where it vanishes as per our  lower boundary condition (red line in Fig. \ref{ux_moy}). Taken jointly, these simulations show that a dynamo magnetic field 
penetrating below the convective zone is able to prevent the viscous diffusion 
of the differential rotation. In this range of magnetic field amplitudes, 
the outcome of our simulations is very similar to that obtained by \citetalias{FD2001} under a different formulation of the upper magnetic boundary condition.

Since the magnetic field amplitude, the effective viscosity and the ohmic diffusivity in the tachocline are not well constrained by observations, 
we also explored a wide range of values for these parameters, to chart the region of our model's parameter space within which tachocline
confinement can be achieved by an oscillating dynamo field penetrating
the radiative zone from above. 

Dimensional analysis yields a first quantitative
estimate 
of the conditions under which this kind of confinement is possible. Specifically,
we construct the dimensionless ratio of the two terms on the
RHS of 
Eq.~\ref{system}.
The Lorentz force (second term) must be stronger than viscous diffusion transport (first term) to suppress the inward spreading and confine the tachocline. This leads to the condition

\begin{equation}
\label{eqn:gamma}
\Gamma = 2\beta \frac{B_0^2}{\rho U_0 \omega/k}\frac{\eta }{\nu} > 1  ~.
\end{equation}

\noindent
Our numerical
results support this dimensional analysis.
Simulations run with $\Gamma<1$ lead to an inexorable inward spreading of the differential rotation. 
A transition takes place as $\Gamma$ approaches and exceeds unity, as the
dynamical effect of the magnetic field becomes important.
Our results indicate that the differential rotation is suppressed by the dynamo field when $\Gamma \gtrsim 3$. 

\subsection{Effect of the magnetic Prandtl number}
\label{subsec:prm}

As we can see in the definition of $\Gamma$ (see Eq. \ref{eqn:gamma}), the confinement of the tachocline depends on the magnetic Prandtl number ($Pr_m = \nu/\eta$). Hence, we explored different viscosities ($\nu$) and ohmic diffusivities ($\eta$) to asses under which conditions the confinement of a viscous tachocline is possible. As expected, we find that high ohmic diffusivities and low viscosities ($Pr_m \ll 1$) lead to an easier confinement of the tachocline. Thus, for a given magnetic field amplitude, there is a critical value of ohmic diffusivity, depending on viscosity, below which it is not possible to stop the inward viscous spreading of the latitudinal differential rotation imposed at the top of the tachocline. As the magnetic field amplitude gets more substantial, this value decreases. 

Starting from our reference model (with $Pr_m=1$), as we decrease the magnetic Prandtl number towards its solar value ($Pr_m=3\times 10^{-2}$, e.g. \citealp{Brun2006}), the confinement is easier and a weaker magnetic field is required. By testing for different magnetic field amplitudes and by evaluating Eq. \ref{eqn:gamma} with $\Gamma=3$, we find that, for the solar magnetic Prandtl number, the critical magnetic field amplitude is of about $70\,G$, a value that is quite inferior to the estimated magnetic field amplitude at the base of the convective zone. Hence, we find that a tachocline subject to viscous diffusion is easily confined by a cyclic dynamo field penetrating below the convective envelope. 

\subsection{Diffusive vs anti-diffusive turbulent viscosity}
\label{subsec:diff_vs_antidif}

To consider the effect of anisotropic turbulence, we decoupled the values of the horizontal and the vertical viscosities. Eq. \ref{system} then becomes

\begin{equation}
\frac{\partial{u}}{\partial{\tau}}=\bigg(\frac{\nu_V}{\eta}\bigg)\frac{\partial^2{u}}{\partial{\xi}^2}-\bigg(\frac{\nu_H}{\eta}\bigg)(k\delta)^2u+C_Lab ~,
\label{u_ani}
\end{equation}

\noindent
where $\nu_V$ stands for the vertical viscosity, while $\nu_H$ represents the horizontal viscosity. We also tested our model with horizontal anti-diffusion, hence with negative horizontal viscosity ($\nu_H=-\nu_V$). Analysis of our simulation results reveals that the effect of vertical viscous transport largely dominates over viscous horizontal transport. This is due to the fact that radial gradients in velocity are much greater than latitudinal gradients in the tachocline. 
 
Vertical diffusive transport works against the confinement of the tachocline. We consider here cases without it ($\nu_V=0$) to determine for which paramater settings the effects of horizontal transport are significant over the Lorentz force. Dimensional analysis of those two terms leads to the following quantity:

\begin{equation}
\label{gamma_hor}
\Gamma_H = \beta\frac{ B_0^2}{\rho U_0 \left|\nu_H \right| k} ~.
\end{equation}

\noindent
For $\Gamma_H>1$, the Lorentz force dominates, while if $\Gamma_H<1$, horizontal diffusion prevails. In the case where $\nu_V=0$ and $\nu_H > 0$, horizontal diffusion helps to confine the tachocline, just as the magnetic field, and the tachocline will always be confined. In the anti-diffusive case ($\nu_H < 0$), it acts in a way to prevent the confinement. Therefore, if the magnetic field is not strong enough or if the negative viscosity is too great, the horizontal diffusion dominates over the Lorentz force, leading to a tachocline that cannot be confined. 

For our reference parameters (with $B_0=1375\,G$, $\eta=7\times 10^9 \, cm^2/s$ and $\nu_H=-7 \times 10^9 \, cm^2/s$), if there is no vertical viscosity, the evolution of the velocity field is dominated by the Lorentz force. Hence, in those system settings, whether the horizontal viscous transport is diffusive or anti-diffusive has a negligeable impact on the results. For our reference model, with $\nu_V=0$ and $\nu_H=-7 \times 10^9 \, cm^2/s$, we find that the critical magnetic field is $B_0=10 \,G$. For a weaker magnetic field, horizontal diffusion amplifies any perturbation of the velocity field throughout the domain and the magnetic confinement of the tachocline fails. Thus, for an anti-diffusive viscosity of $\nu_H=-7 \times 10^9 \, cm^2/s$, the magnetic field needed for the tachocline to be confined is very low. 

We now consider the expected magnetic field amplitude ($B_0=1375\, G$) and tested our model with enhanced negative viscosities. We find that the tachocline is not confined only when $|\nu_H| \gtrsim 10^{12} \, cm^2/s$. For such a strong negative viscosity, the Lorentz force is not strong enough to prevent the inexorable growth of the velocity caused by anti-diffusion. For intermediate viscosities of about $\nu_H=-10^{10-12} \, cm^2/s$, the horizontal viscous transport starts to play a greater role in the evolution of the velocity field. As the viscosity becomes more important, horizontal diffusion causes strong oscillations on the velocity profile. 

We conclude that if the horizontal diffusivity is of the same amplitude as the vertical diffusion, its impact on the evolution of the magnetic field is negligible (equivalently, one may set $\nu_H \sim 0$). Hence, neglecting the horizontal diffusion leads to the same results as those discussed in Sect. \ref{subsec:results}. Also, if turbulence is anisotropic and vertical diffusion is negligible compared to horizontal anti-diffusion ($\nu_V \sim 0$), the Lorentz force is likely to dominate over horizontal diffusion. Indeed, the effects of horizontal diffusion start to be significant only for very weak magnetic field amplitudes or strong diffusivities, extreme cases that are not realistic while considering solar internal conditions. In that case, whether the tachocline is subject to diffusion or anti-diffusion in the horizontal direction has no effect on the conclusions. Finally, if turbulence is anisotropic in a way that both horizontal anti-diffusion and vertical diffusion have an impact on the evolving shear ($| \nu_H  | > \nu_V$, but $\nu_V$ is not negligible), both effects act together to spread the tachocline downwards. This situation is the most difficult to confine, but we observe that even for very strongly enhanced (anti) viscosities ($| \nu_H  |, \nu_V < 10^{12} \,cm^2/s$) and a moderate magnetic field ($B_0 \sim 10^3 \, G$), the confinement of the tachocline by the cyclic magnetic field can still be realized.

\section{Magnetic tachocline subject to radiative spreading}
\label{sec:hyper}

\subsection{1D model with radiative spreading}
\label{subsec:justhyper}

The numerical results reported upon in
Sect. \ref{sec:magtach} were based on the assumption 
that the hydrodynamical contribution to angular momentum transport 
was only due to turbulent viscosity. However, \cite{Tobias2007} show that in the presence of magnetic fields, the turbulent transport might not have 
an impact on the transport of angular momentum, owing to the tendency of Maxwell
stresses to offset Reynolds (turbulent) stresses.

In this case, the dominant contribution would more likely be radiative spreading, as proposed by \citetalias{Spiegel1992}. The radiative spreading can be understood as follows. Because the thermal relaxation time is short in the tachocline, meridional flows are controlled by thermal diffusion. The advective angular momentum transport by the meridional flow is then also controlled by thermal diffusion, and \citetalias{Spiegel1992} showed it could be interpreted as a 1D (radial) hyperdiffusive term in the evolution equation for the differential rotation (see Sect. 4 in \citetalias{Spiegel1992}). Because radiative spreading sets in when the thermal diffusion time is short compared to the viscous time ($Pr \ll 1$), we ignored viscous diffusion altogether in the following analysis, setting the effective viscosity to zero, hence replacing Eq. \ref{system} with  

\begin{equation}
\frac{\partial{u}}{\partial{\tau}}=C_Lab-\rm{Fr} \space \bigg(\frac{\kappa}{\eta}\bigg)\bigg(\frac{R}{\delta}\bigg)^2\frac{\partial^4{u}}{\partial{\xi}^4} ~.
\label{u_hyp}
\end{equation}

\noindent
In Eq. \ref{u_hyp}, $\rm{Fr}$ is the Froude number, the value of which is
set by the stratification
\begin{equation}
\rm{Fr}=(2\Omega/N)^2 ~, 
\qquad
\text{with}
\quad
 N^2=g\bigg( \frac{1}{\gamma}\frac{d (\text{ln} P)}{dr} - \frac{d( \text{ln} \rho)}{dr} \bigg) ~,
\end{equation}

\noindent
where $g$ is the gravitational acceleration and $\gamma=(\partial$ln$ P / \partial $ln$ \rho)_{ad}$ is the adiabatic exponent. Dimensional analysis indicates that the ratio of the viscous term (in Eq. \ref{system}) over the hyperdiffusive term (in Eq. \ref{u_hyp}) scales as $Pr/Fr(\delta/R)^2 \sim 10^{-2}$ for the solar case, confirming that viscous diffusion can be neglected over radiative spreading. A tachocline subject to radiative spreading will be harder to confine since thermal diffusivities greater than ohmic diffusivities (as in the tachocline) tend to favor the inward propagation of the differential rotation. 

\subsection{Choice of parameters}
\label{subsec:param_rs}

The inclusion of an hyperdiffusive term in Eq. \ref{u_hyp} introduces
yet another parameter, the thermal diffusivity $\kappa$, which controls how dominant the radiative spreading is. 
We used the same domain size, horizontal wavenumber, period, density and differential rotation amplitude as in Sect. \ref{sec:magtach}. We explore a range of values for the ohmic and thermal diffusivities, as well
as various magnetic field amplitudes, as noted in Table \ref{table_sect3}. 
We retain the same ratio of toroidal-to-poloidal magnetic field amplitudes ($\beta=10$). 
For simplicity, we choose to fix the Froude number to $2 \times 10^{-6}$ (e.g. \citealp{Brun2006}). We consider it constant over the whole studied domain and we do not take into account the evolution of the stratification over the Sun's lifetime. This is compensated by the variation of other parameters as the thermal diffusivity. 
To begin with, as in Sect. \ref{sec:magtach}, poloidal and toroidal magnetic field amplitudes are set respectively to $1375\,G$ and $13750\,G$ and the ohmic diffusivity is fixed at $\eta=7\times10^9 \, cm^2/s$. 
We choose to compare a case with thermal diffusion equal to the viscous diffusion in Sect. \ref{sec:magtach}, hence, with a thermal diffusivity of $\kappa=7\times10^9 \, cm^2/s$.

%
%

\begin{table}
\centering
\caption{Numerical values for the parameters involved in the model including radiative spreading.}
\begin{tabular}{ccc}
\hline
\hline
 \footnotesize{Parameter} & \footnotesize{Symbol} & \footnotesize{Numerical value} 
 \\[0.5ex]

\hline
\footnotesize{Ohmic diffusivity} 				    & \footnotesize{$\eta$}			 	& \footnotesize{$10^{5}-10^{10}\,cm^2/s$}  \\
\footnotesize{Froude number}		 				    & \footnotesize{$\rm{Fr}$}			 	& \footnotesize{$2 \times 10^{-6}$}	    \\
\footnotesize{Thermal diffusivity}			 		& \footnotesize{$\kappa$}			& \footnotesize{$10^5-10^{12}\,cm^2/s$}  \\	
 \hline
\end{tabular}
\label{table_sect3}
\end{table}

\subsection{Results}
\label{subsec:resultshyper}

\begin{figure}[h]
\centering
\includegraphics[width=0.99\linewidth]{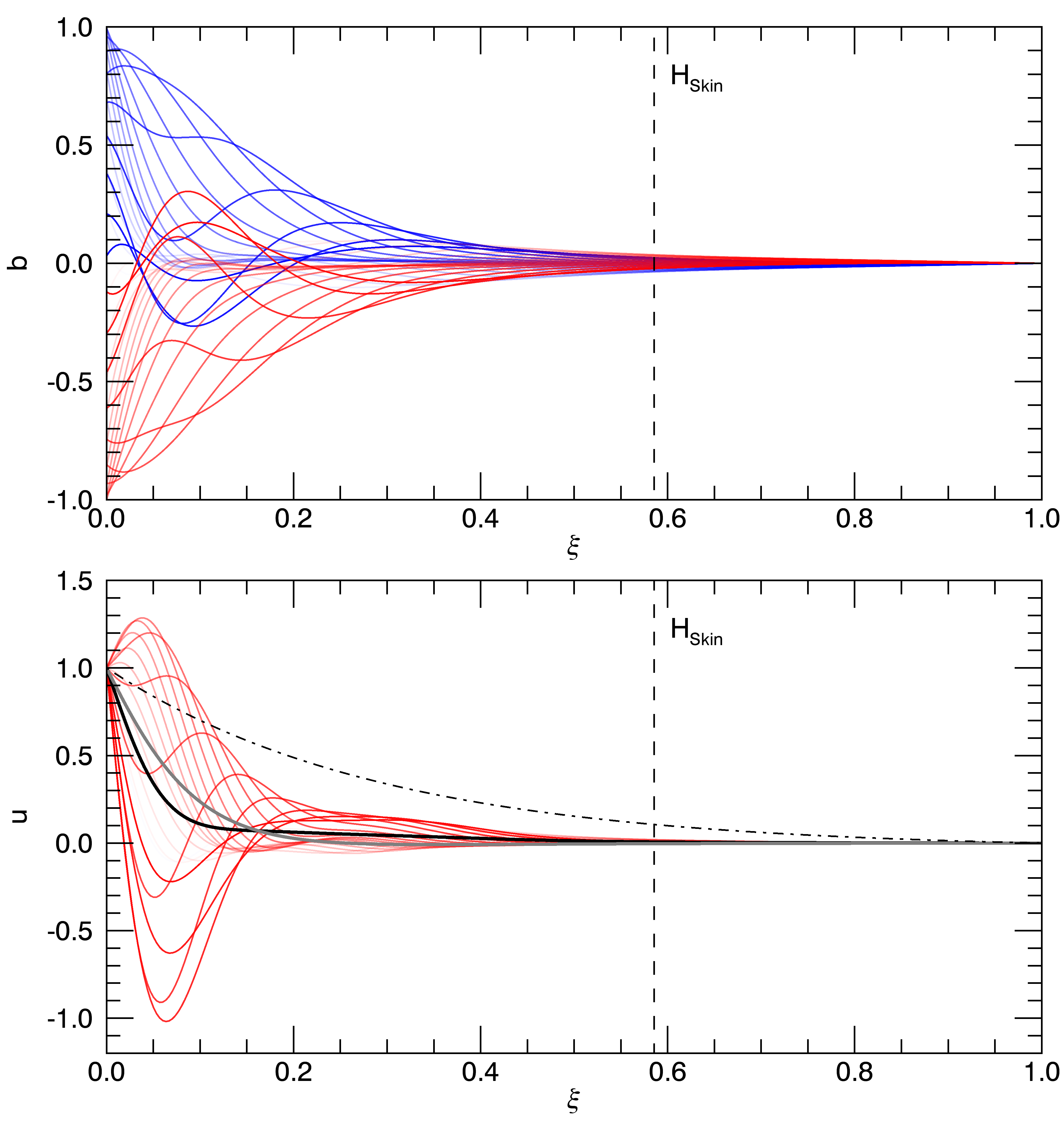}
\caption[Temporal evolution of $U_x$ (with radiative spreading)]{\label{evol_ux_sr} \footnotesize{(A) Temporal evolution of the toroidal magnetic field ($b$) and (B) the azimuthal velocity field amplitude ($u$) as a function of depth, for one cycle, for a fully relaxed simulation ($\tau=100$). For the velocity field, only half a cycle is shown as profiles are superimposed for the second half. Lines become darker as time proceeds. The skin depth (dashed line and Eq. \ref{eqn:skin}) and the initial condition for $u$ (dash-dotted line) are shown. 
The black line is the azimuthal velocity amplitude averaged over one cycle. This field is subject to radiative spreading and the Lorentz force. As in Sect. \ref{sec:magtach}, poloidal and toroidal magnetic field amplitudes are respectively $1375\,G$ and $13750\,G$ and the phase shift between both magnetic fields is $\phi=\pi/2$. The ohmic diffusivity is $\eta=7\times10^9 \, cm^2/s$ and thermal diffusivity is $\kappa=7\times10^9 \, cm^2/s$. The mean velocity amplitude for the viscous case in Fig. \ref{evol_visc}}C is shown in grey.}
\end{figure}

Figs. \ref{evol_ux_sr}A-\ref{evol_ux_sr}B show the temporal evolution of the toroidal magnetic field and the azimuthal velocity amplitude as a function of depth for one cycle, as in Figs. \ref{evol_visc}B-\ref{evol_visc}C. The velocity field has a period of half the period of the magnetic fields. 
As in the viscous case, the simulation reached a stationary state after a few diffusion times.
One first difference worth noting between both cases is that, with radiative spreading now acting, the velocity field is subject to greater oscillations owing to the inclusion of the hyperdiffusive term. We can also see that, at some points in
the cycle, $u$ is greater than $1$: for some instants, the shear imposed at the top of the domain is amplified by the dynamo field. 
Despite these differences, we see in Fig. \ref{evol_ux_sr}B that the velocity profiles averaged over all cycle phases for a tachocline subject to viscous diffusion (in grey, as in Fig. \ref{evol_visc}C) or radiative spreading (in black) are similar. In the case studied here the latitudinal shear vanishes over a depth that is much smaller than the extent of our solution domain, which implies that the tachocline is confined. We see that the confinement is more efficient since the black curve is narrower than the grey curve.

\begin{figure}[h]
\centering
\includegraphics[width=0.99\linewidth]{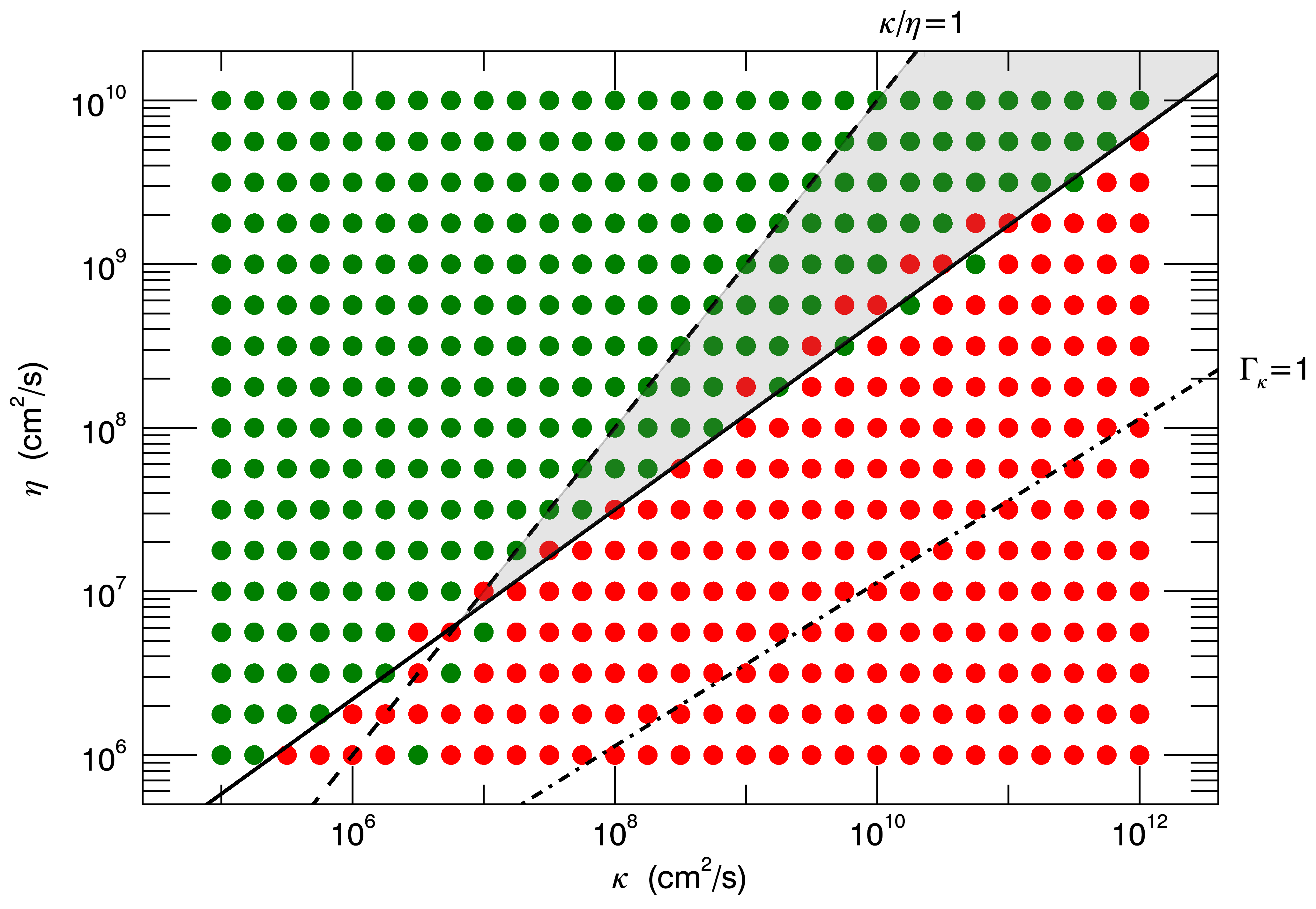}
\caption[Parameters space for $B_0=5000 \, G$]{\label{param}  \footnotesize{Results of simulations for different values of ohmic and thermal diffusivities ($\eta = 10^6 - 10^{10} \, cm^2/s$, $\kappa = 10^5 - 10^{12} \, cm^2/s$) and for a poloidal magnetic field amplitude of $5000 \, G$. Green dot are simulations for which the tachocline is confined, while red dots are not confined. The solid and dashed lines indicate respectively the approximated transition between confined and unconfined tachoclines and the simulations where $\kappa/\eta=1$. The dash-dotted line labels $\Gamma_{\kappa}=1$ (see Eq. \ref{eqn:gamma_kappa})}} 
\end{figure}

Fig. \ref{param} shows the results of different simulations, carried out
with $B_0=5000\,G$ and a large interval of ohmic and thermal diffusivities. We even considered here cases with $\kappa < \eta$, for completeness (we will come back to this point below). Simulations performed with a set of parameters which leads to a confinement of the tachocline are represented by green dots, while red dots are simulations where the magnetic confinement fails. We consider the tachocline to be confined if the three following criteria are fulfilled:

\begin{itemize}
\item[$\bullet$] in the studied domain, the second derivative of the mean velocity ($\bar{u}$) must be zero at least twice;
\item[$\bullet$] in the lower part of the tachocline ($\xi>0.75$), the mean velocity must be small ($ \bar{u} < 0.018 $);
\item[$\bullet$]  in the lower part of the tachocline ($\xi>0.75$), the first derivative of the mean velocity must be low ($ \partial_{\xi} \bar{u} < 0.15 $). 
\end{itemize}

\noindent
The first and last conditions ensure that a real damping takes place in the studied domain, while the second condition makes sure that no spurious boundaries effects control the solution. 
Those criteria are somewhat arbitrary, but they determine unambiguously confined vs unconfined states and they are used in a consistent way for the whole set of simulations presented in this work. A modification of these conditions could change slightly the confinement state near the transition, but would not alter the overall picture. 

We see a clear tendency; for strong ohmic diffusion and weak thermal diffusion, the interaction between the fields leads to the suppression of differential rotation. On the other hand, for weak ohmic diffusion and strong thermal diffusion, there is no confinement. This tendency is robust for different magnetic field amplitudes. 

As in Sect. \ref{sec:magtach}, we compared the terms that influence the evolution of the velocity by dimensional analysis.
For the tachocline to be confined, the Lorentz force must be stronger than the hyperdiffusion. Hence, we must have

\begin{equation}
\label{eqn:gamma_kappa}
\Gamma_{\kappa} = \frac{4 \beta}{\rm{Fr} } \frac{B_0^2}{\rho U_0 \omega/k} \frac{\eta}{\kappa} \frac{\eta}{\omega R^2} > 1 ~.
\end{equation}

\noindent
The relation between $\eta$ and $\kappa$ for $\Gamma_{\kappa}=1$, is shown in Fig. \ref{param}. We can see that, as expected, when $\Gamma_{\kappa}<1$, there is no confinement of the tachocline. For $\Gamma_{\kappa}>1$, there is a transition phase, where the Lorentz force begins to play a significant role. Values $\Gamma_{\kappa} \gg 1$ are now required for the magnetic field to be able to confine the tachocline and our results indicate that the differential rotation is suppressed by the dynamo field when $\Gamma_{\kappa} \gtrsim 500$. 
In this transition phase, since the Lorentz force and the radiative spreading both have a significant influence, the system tends toward an equilibrium state where the velocity as a function of the depth is a parabola. Hence, for a given magnetic field amplitude and thermal diffusivity, a large ohmic diffusivity is needed for the system to overcome this transition phase and for the velocity field to be suppressed. 

\begin{figure}[h]
\centering
\includegraphics[width=0.99\linewidth]{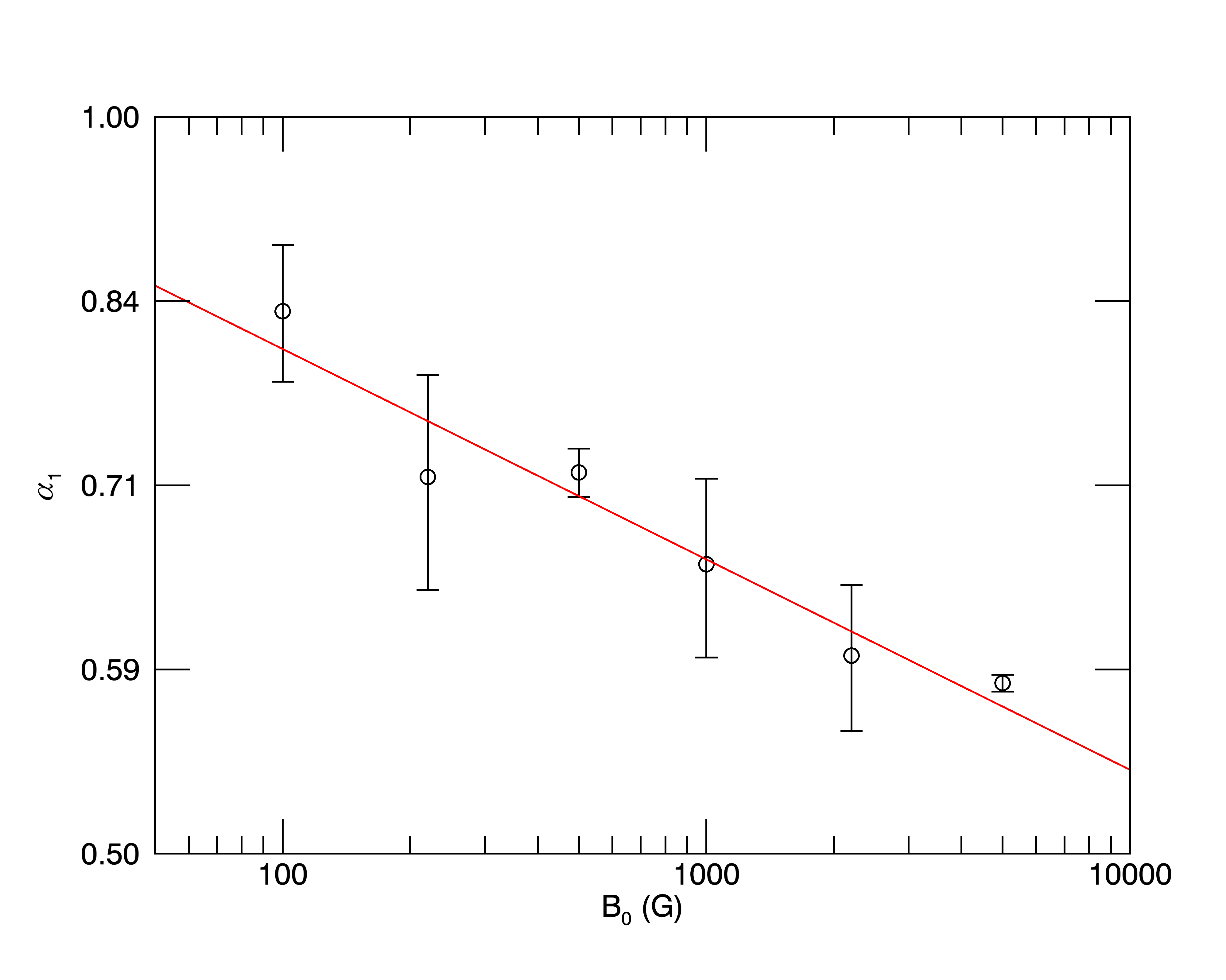}
\caption[$\alpha_1$ as a function of magnetic field amplitude]{\label{a_vs_b0} \footnotesize{ $\alpha_1$ term in Eq. \ref{eqn:alpha_vs_b0} as a function of $B_0$, leading to $\alpha_2=-0.085\pm 0.012$.}} 
\end{figure}

The relationship between the critical values of ohmic and thermal diffusivities ($\eta_{crit}$, $\kappa_{crit}$) separating confined vs not confined states follows a power law (Eq. \ref{eqn:alpha_vs_b0}), with an index ($\alpha_1$) that also scales approximately 
as a power law for the magnetic field amplitude, as seen in Fig. \ref{a_vs_b0}. 

\begin{equation}
\begin{split}
\begin{gathered}
\label{eqn:alpha_vs_b0}
\eta_{crit}\propto\kappa_{crit}^{\alpha_1} \quad ~;\quad  \alpha_1 \propto B_0^{\alpha_2}\quad  ~; \quad  \alpha_2=-0.085 \pm 0.012
\end{gathered}
\end{split}
\end{equation}

\noindent
The values of $\alpha_1$ were obtained by fitting the confinement transition for different values of $B_0$ (see Fig. \ref{param}, for $B_0=5000\, G$). We see that for larger magnetic field, $\alpha_1$ becomes smaller, which implies that the confinement of the tachocline is less sensitive to thermal diffusion.

In the solar tachocline, the microscopic thermal diffusivity is $\kappa \approx 10^7\,cm^2/s$ and the microscopic ohmic diffusivity is $\eta \approx 10^3\,cm^2/s$. 
(see, e.g., \citealp{Brun2006}).
For such a low ohmic diffusivity, the dynamo magnetic field penetrates only
an extremely thin layer at the bottom of the convective zone, which is not reasonable given that convective overshoot likely penetrates to a much larger depth. 
Moreover, by extrapolating Fig. \ref{param}, we can conclude that for these microscopic parameters values, the tachocline could not be confined by an oscillatory magnetic field. However, for a turbulent tachocline, the turbulent ohmic diffusivity would be a few orders of magnitude stronger. \citetalias{FD2001} concluded that this value should be of about $\eta=10^9 - 10^{10}\,cm^2/s$ for the tachocline to be confined by an oscillatory magnetic field. 
According to Fig. \ref{param}, we find that, for a thermal diffusivity of $\kappa=10^7\,cm^2/s$ and a poloidal magnetic field amplitude of $B_0=5000 \, G$, an ohmic diffusivity of $\eta=10^7 - 10^8\,cm^2/s$ would already be sufficient to stop the inward spreading of differential rotation. 

\begin{figure}[h]
\centering
\includegraphics[width=0.99\linewidth]{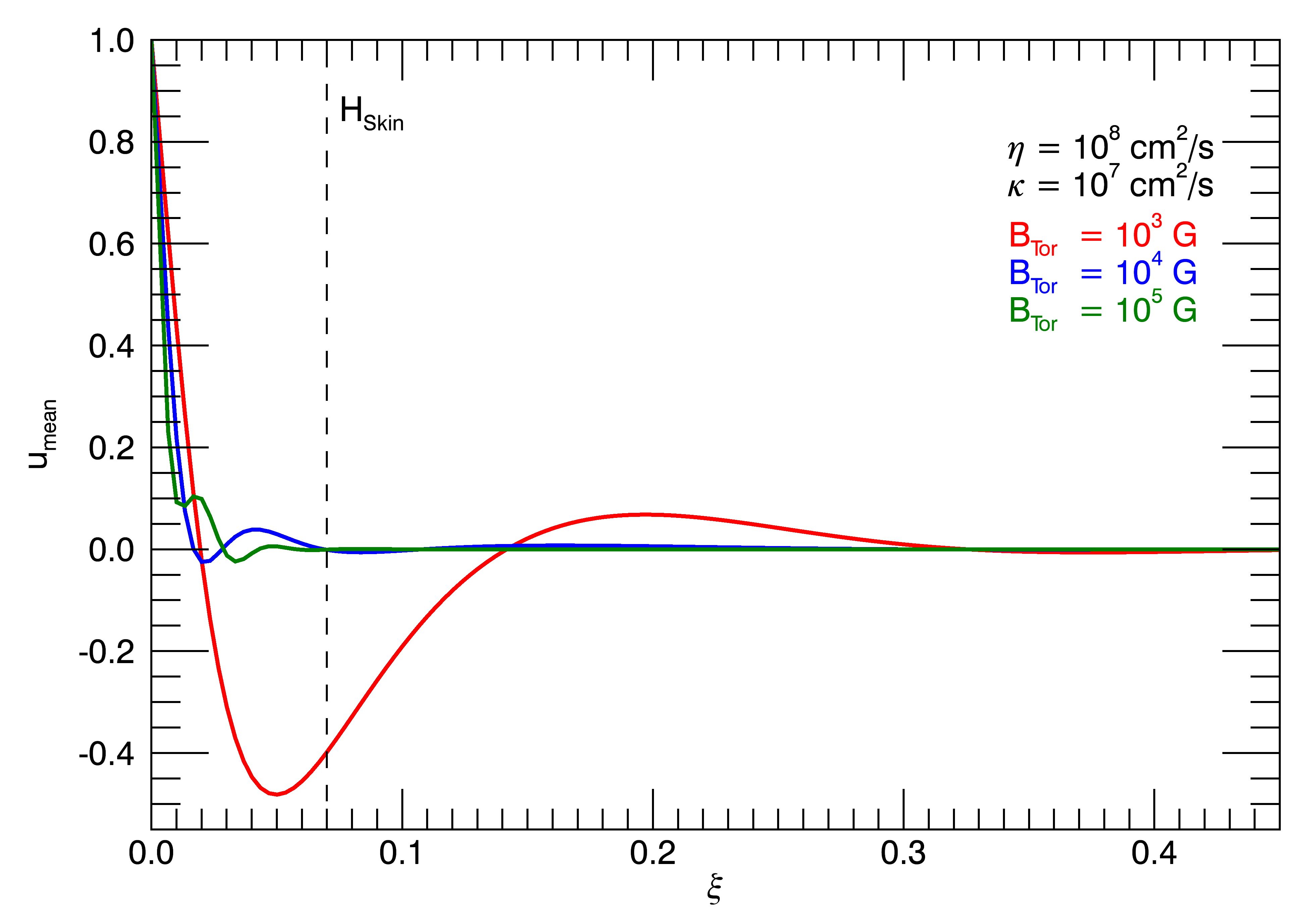}
\caption[Mean azimuthal velocity for different magnetic field amplitudes]{\label{diff_b0} \footnotesize{ Mean normalized azimuthal velocity for a turbulent solar tachocline ($\eta_{turb} = 10^8 cm^2/s$, $\kappa = 10^7 cm^2/s$) for different magnetic field amplitudes. Greater magnetic fields lead to an easier confinement.}} 
\end{figure}

Figure \ref{diff_b0} shows the mean azimuthal velocity for cases with a solar thermal diffusivity ($\kappa=10^7 \, cm^2/s$) and a turbulent ohmic diffusivity ($\eta=10^8 \, cm^2/s$) with different magnetic field amplitudes. We see that, as the magnetic field amplitude increases, the mean velocity is suppressed over a smaller depth. 

However, our results also show that the amplitude of the oscillations over one period increases drastically as the magnetic field amplitude gets more substantial. Indeed, while for a toroidal magnetic field amplitude of $10^3 \, G$, the velocity variation over one magnetic cycle is only of a few percents, for a toroidal magnetic field amplitude of $10^5 \, G$, $u$ oscillates between $2 \, U_0$ and $-3 \, U_0$, a variation of about $500\%$.

As seen in Fig. \ref{evol_ux_sr}, our results show strong oscillations in the velocity over the magnetic field cycle. Evidence for short timescale oscillations at the base on the convective zone was detected \citep{Howe2011}, but they do not appear robust over several solar cycles. The shear variations in our simulations take place within a relatively thin layer, which is equivalent to the skin depth (see Eq. \ref{eqn:skin}). This depth is larger for higher ohmic diffusivity. The spatial resolution limit of helioseismic observations is of about $5\%$ of the solar radius \citep{Howe2009} which leads to a reasonable upper limit on the ohmic diffusivity of about $5 \times 10^{10} \, cm^2/s$. Hence, higher ohmic diffusivities were excluded from the plausible scenario for the solar tachocline in this study.

We also performed experiments with both viscous diffusion and radiative spreading. We find that for $Pr$ lower or equal to $1$, radiative spreading dominates the evolution of the velocity field. For $Pr>1$, viscosity begins to dominate and the purely viscous cases presented in Sect. \ref{sec:magtach} are typically recovered when $Pr>10$. For cases where both angular momentum transport processes have an impact, the velocity field oscillations mentioned above are only slightly attenuated by the additional viscous contribution.

If turbulence carries heat diffusively in the tachocline, we can consider a thermal diffusivity greater then the atomic value. In this case, since, microscopically, thermal diffusivity is greater than ohmic diffusivity ($\kappa > \eta$), the parameters set of our model should, a priori, respect this hierarchy to be solar-like (this aspect, though, should be confirmed with dedicated numerical simulations of turbulence in the tachocline). 
In Fig. \ref{param}, the dashed line correspond to $\kappa/\eta=1$. 
To obtain a confined tachocline with a thermal diffusivity greater than the ohmic diffusivity, the set of parameters needs to be within the shaded region delimited by the solid and the dashed lines in Fig. \ref{param}. For stronger magnetic fields, this region broadens and a greater number of parameter sets $\{ \kappa, \eta \}$ with $\kappa/\eta > 1$ leads to a confined tachocline.

\section{Discussion and conclusion}
\label{sec:conclu}

The physical processes responsible for maintaining the small thickness of
the solar tachocline have been debated ever since the original solution of
\cite{Spiegel1992}.
Many possible explanations have been put forth over the last two decades, 
but no definitive conclusion, let alone concensus, has been reached.
Among the different hypotheses proposed, \citetalias{FD2001} suggested a 
fast tachocline scenario, in the context of which an oscillatory poloidal magnetic field generated by dynamo action within the convective envelope
penetrates into the underlying radiative core, providing magnetic stresses
that suppress latitudinal differential rotation in the tachocline. The strength of this scenario is that we know for a fact that the Sun possesses an 11-year cyclic field, that has to penetrate, to some extend, the upper part of the tachocline. In their original model, \citetalias{FD2001} imposed a differential rotation profile and a poloidal magnetic field at the base of the convection zone. 
At first, the poloidal magnetic field diffuses inward under the influence
of ohmic diffusion, while the differential rotation spreads inward as a result of viscous diffusion. 
This leads to the buildup of a 
shear, which induces a toroidal magnetic field. On to this is
associated a large-scale Lorentz force which can prevent the burrowing of
differential rotation through efficient equator-to-pole transport
of angular momentum by magnetic stresses.  

We tested \citetalias{FD2001}'s model by applying a more general
boundary condition for the toroidal magnetic field than the zero-field
condition they used.
In our model, a toroidal magnetic component varying periodically is also imposed
at the top of the radiative zone, oscillating with a set
phase lag with respect to the poloidal magnetic component. We follow
\citetalias{FD2001} in imposing a fixed
latitudinal shear at the top of the domain.
Our results show that such a magnetic field is able to prevent the viscous diffusion of the shear, and that the outcome depends only moderately on the phase difference imposed between the two oscillating magnetic components. 
To prevent the spreading of the differential rotation, the medium needs to be turbulent, with enhanced viscosity and ohmic diffusivity of about $\nu \approx \eta \approx 10^8 \, cm^2/s$. These results are in agreement with those obtained by \citetalias{FD2001}.

The turbulent transport of angular momentum
in the tachocline is another controversial subject. Indeed, in the horizontal direction, it could be diffusive (\citealp{Miesch2003}, \citealp{Spiegel1992}, \citealp{Kim2007}) or anti-diffusive \citep{Gough1998}. 
It could also be neither, since in the presence of magnetic fields, Reynolds stresses tend to cancel Maxwell stresses \citep{Tobias2007}. In this case, even in the
presence of strong turbulence, the transport of angular momentum could still
remain dominated by radiative spreading, albeit the fully magnetized case of radiative spreading still needs to be studied in details. 

To complement \citetalias{FD2001}'s work, we tested the fast tachocline scenario with different turbulent transport mechanisms. We found that whether the viscous horizontal transport is diffusive or anti-diffusive has no significant
impact on the results since the vertical transport is dominant, except in extreme and physically unrealistic parameter regimes.  

Our results show that a viscous tachocline with a turbulent viscosity of $\nu=10^7 \, cm^2/s$ and a turbulent ohmic diffusivity of $\eta=10^8 \, cm^2/s$ can be confined by a dynamo magnetic field ($B_{Tor}=5\times 10^4 \, G$ and $B_{Pol}=5\times 10^3 \, G$) that penetrates below the convective envelope, whether the horizontal turbulent transport is diffusive or anti-diffusive. Many other sets of parameters also lead to a suppression of the differential rotation in the tachocline. 

We then consider a tachocline subject to radiative spreading, as the dominant transport process of angular momentum. We find that a dynamo field is able to stop the burrowing of the differential rotation in the radiative zone for certain combinations of magnetic field amplitudes and thermal and ohmic diffusivities. Our results also show that this confinement is easier for greater magnetic field amplitudes and ohmic diffusivities and for lower thermal diffusivities, as expected. 
These results are constrained by the physical conditions in the solar tachocline.
The microscopic solar thermal diffusivity in the tachocline, being about $\kappa \approx 10^7 \, cm^2/s$,
implies a minimal ohmic diffusivity of about $\eta \approx 10^{7-8} \, cm^2/s$. 

In this scenario, oscillations of the velocity field with the magnetic cycle are inevitable. Given the current spatial resolution limit of helioseismic rotational inversion \citep{Howe2009}, our results restrict the ohmic diffusivity to a value $\eta \lesssim \times 10^{10} \, cm^2/s$ (for higher $\eta$, the associated oscillations of the velocity field would a priori be detectable in helioseismic inversions).

We also explored cases with both turbulent viscous transport and radiative spreading. For Prandtl numbers ($Pr$) lower or equal to $1$, we retrieve solutions for a tachocline only subject to radiative spreading. The impact of viscosity dominates the evolution of the velocity field for $Pr>10$. There is a transition during which viscosity slightly alters the velocity field amplitude, but this does not modify our overall conclusion about the confinement of the tachocline.

Our results show that a tachocline subject to radiative spreading with a turbulent thermal diffusivity of $\kappa=10^9 \, cm^2/s$ and a turbulent ohmic diffusivity of $\eta=10^8 \, cm^2/s$ can be contained by the penetration of a dynamo magnetic field ($B_{Tor}=5\times 10^4 \, G$ and $B_{pol}=5\times 10^3 \, G$) into the radiative zone. Hence, tachocline confinement remains possible over a relatively wide range of parameter values, including these so-called solar values. \\
\bigskip

\noindent
\footnotesize{
\textit{Acknowledgements.}
We dedicate this work to J.P. Zahn, who sadly passed away while we were conducting this study with him. His work and enthusiasm about the solar tachocline always have been a source of great inspiration for us.
We thank our colleagues E. Forg\'{a}cs-Dajka and K. Petrovay for useful discussion. R. Barnab\'e was supported by a graduate fellowship from Le Fonds de Recherche du Qu\'ebec - Nature et Technologies (FRQNT). A. Strugarek acknowledges support from the Canadian Institute of Theoretical Astrophysics (National Fellow). This work was also supported by Canada's Natural Sciences and Engineering Research Council. A.S. Brun acknowledges funding by INSU/PNST, CNES SolarOrbiter, PLATO and GOLF grants.
}


\end{document}